\documentclass[11pt,letterpaper]{article}
\usepackage{jheppub}
\usepackage{mathrsfs}
\usepackage{amsmath,amssymb,graphicx}
\usepackage {amssymb}
\usepackage{color}
\newcommand{\nc}{\newcommand}
\nc{\bea}{\begin{eqnarray}} \nc{\eea}{\end{eqnarray}}
\nc{\be}{\begin{equation}} \nc{\ee}{\end{equation}}

\newcommand\h{\hat}

\input{epsf.sty}

\begin{document}

\title{Reconstruction of a Nonminimal Coupling Theory with Scale-invariant Power Spectrum}

\author{Taotao Qiu$^{1,2}$}
\emailAdd{xsjqiu@gmail.com}
\emailAdd{qiutt@ntu.edu.tw}

\vspace{16mm}

\affiliation{$1$. Leung Center for Cosmology and Particle Astrophysics National Taiwan University, Taipei 106, Taiwan}
\affiliation{$2$. Department of Physics, National Taiwan University, Taipei 10617, Taiwan}

\vspace{16mm}

\date{\today}

\keywords{Nonminimal coupling, Conformal transformation, Inflation, Matter-contraction}


\abstract{
A nonminimal coupling single scalar field theory, when transformed from Jordan frame to Einstein frame, can act like a minimal coupling one. Making use of this property, we investigate how a nonminimal coupling theory with scale-invariant power spectrum could be reconstructed from its minimal coupling counterpart, which can be applied in the early universe. Thanks to the coupling to gravity, the equation of state of our universe for a scale-invariant power spectrum can be relaxed, and the relation between the parameters in the action can be obtained. This approach also provides a means to address the Big-Bang puzzles and anisotropy problem in the nonminimal coupling model within Jordan frame. Due to the equivalence between the two frames, one may be able to find models that are free of the horizon, flatness, singularity as well as anisotropy problems.}

\maketitle

\section{Introduction}

It is well known that although Einstein's general relativity (GR) has been extremely successful in describing our universe, possibilities of extensions of GR has not been ruled out yet. Gravity can be modified in various ways, while one of the simplest ways might be to couple a scalar field $\phi$ nonminimally to the Ricci scalar $R$, with terms such as $F(\phi)R$ \cite{Abbott:1981rg,Futamase:1987ua}. Besides its motivation from fundamental theories, e.g. the low-energy effective string theory, where $\phi$ can act as a dilaton \cite{Gasperini:1992em}, phenomenologically such kind of extension has been widely applied to cosmological models in the early universe, for it can not only solve various problems, but also present phenomenological predictions. For example, in the inflation scenario (\cite{Guth:1980zm,Albrecht:1982wi,Linde:1983gd}, also see \cite{Starobinsky:1985ww} for reviews) where $\phi$ acts as an inflaton, it has been found that the inflation phase could be more easily obtained compared to the case without nonminimal coupling, and an attractor solution is also available \cite{Futamase:1987ua}. Furthermore, nonminimal coupling terms in inflation may also give rise to corrections on power spectrum of primordial perturbations \cite{Salopek:1988qh} and non-Gaussianities \cite{Qiu:2010dk}, as well as a tiny tensor-to-scalar ratio \cite{Komatsu:1997hv}, which can be used to fit the data or to constrain the parameters. Inflation with nonminimal coupling can also provide other applications, such as the realization of warm inflation \cite{Bellini:2002zr}, or the avoidance of the so-called $\eta$-problem \cite{Copeland:1994vg} in the framework of string theory \cite{Easson:2009kk} and so on, see \cite{Faraoni:2000gx} and \cite{DeFelice:2010aj} for comprehensive reviews. Besides this, there are also other versions of modified gravity, some of which are proposed quite recently, such as $f(R)$ theories \cite{Bergmann:1968ve}, Lovelock theories \cite{Lovelock:1971yv}, Horava-Lifshitz theories \cite{Horava:2008ih}, teleparallel theories \cite{Unzicker:2005in}, massive gravity \cite{deRham:2010ik}, etc.

Any viable theory must satisfy observational constraints. For theories of the early universe, it must generate the right amount of perturbations so as to conform with the observable cosmic microwave background (CMB) anisotropies \cite{Larson:2010gs} and large scale structure (LSS) \cite{Bernardeau:2001qr}. The power spectrum of these perturbations, which comes from the 2-point correlation function, has to be (nearly) scale-invariant \cite{Larson:2010gs}. These will put on non-trivial constraints on models based on nonminimal coupling theories.

However, when one starts with a nonminimal coupling theory, one is in general free to choose any form of the action, especially its coupling and potential terms. Due to such an arbitrariness, the analytical calculation of the spectrum usually becomes complicated, and thus numerical methods have to be involved. Actually, it is found that if the coupling is linear in $R$, i.e., has the form of $F(\phi)R$, the theory can be transformed to another frame, e.g. the Einstein frame, just by performing a conformal transformation on the metric, namely \cite{Faraoni:1998qx}: \be\label{conformal} \h{g}_{\mu\nu}=\Omega^2g_{\mu\nu}~,\ee where $g_{\mu\nu}$ with and without hat denotes the metric in the Einstein and Jordan frames, respectively. It is well-known that the theories in the two frames are equivalent to each other, namely, they should make the same observable results \cite{Makino:1991sg}, while in the Einstein frame, it looks like a normal scalar theory in GR, since the nonminimal coupling term has been absorbed by the transformation. \footnote{It may not be the case when multi-field or matter get involved due to the reason of ``running units", see \cite{Faraoni:2006fx}.} For this reason, when dealing with the nonminimal coupling theories, people always tend to move to its Einstein frame and calculate everything there. But inversely, given its counterpart in the Einstein frame, which is much simpler to deal with, these properties also provide us another way of obtaining a nonminimal coupling theory with appropriate observational features. Compare with the more conventional ``{\it from theory to phenomenology}" process, this is actually a reversed ``{\it from phenomenology to theory}" process. For example, it is already known that there are at least two scenarios in which a single scalar field theory can give rise to scale-invariant perturbations, namely inflationary expansion \cite{Guth:1980zm,Albrecht:1982wi,Linde:1983gd,Starobinsky:1985ww}, or matter-like contraction \cite{Finelli:2001sr}. If we take them merely as an Einstein frame presentation of some nonminimal coupling theory, then we can go back to find its counterpart form in the Jordan frame. That is the main goal of this paper.

What is the ``Einstein frame" inflation/matter-contraction like in its Jordan frame? It is non-trivial to answer this question, since we don't know {\it a priori} what the form of $\Omega^2$ in Eq. (\ref{conformal}) is, so we perform a reconstruction approach for each of the two cases. Note that reconstruction methods has also been used to obtain $f(R)$ or scalar-tensor theories in the expanding universe \cite{Nojiri:2006be}. By reconstruction, we obtain the relation between the evolution behavior of the universe and the parameters in the action, namely, if we give appropriate parameters to the action and let it evolve as expected, then we can get a scale-invariant power spectrum which fits the observational data. The content of this paper is organized as follows: in Sec. II we first demonstrate the relation between Jordan and Einstein frames. In the background level, we derived the transformations of various variables between the two frames, and more importantly, in the perturbation level, we show that the perturbations are invariant under this conformal transformation, which is the basis of our reconstruction. We also give conditions under which a scale-invariant power spectrum can be obtained in both frames. In Sec. III, we perform our reconstruction of a nonminimal coupling theory with scale-invariant power spectrum, both from inflation and matter-contraction scenarios driven by a minimal coupling field, which can act as its Einstein presentation. Relations between the equation of state and the parameters in the action are obtained. In Sec. IV, we discuss on some other issues of the early universe. Sec. V comes our conclusions.

\section{Jordan and Einstein Frames}
\subsection{background evolution}
In this subsection, we perform the relations of the background variables of a nonminimal coupling theory in its Jordan and Einstein frames, in latter of which it presents like a minimal coupling one. First of all, we begin with the action in Jordan frame: \be\label{actionJ} {\cal S}_J=\int d^4x\sqrt{-g}\Bigl[M_{Pl}^2F(\phi)R-\frac{1}{2}Z(\phi)\partial_\mu\phi\partial^\mu\phi-U(\phi)\Bigr]~,\ee where $F(\phi)$ and $Z(\phi)$ can be arbitrary functions of the field $\phi$ in the Jordan frame, and $U(\phi)$ is the potential. Note that when $F(\phi)=[1-\xi(\phi/M_{Pl})^2]/2$ and $Z(\phi)=1$, it reduces to the first/simplest nonminimal coupling form \cite{Abbott:1981rg,Futamase:1987ua} that has been widely studied in the literature (see e.g. \cite{Salopek:1988qh,Faraoni:2000gx,Uzan:1999ch,Bassett:1997az,Bezrukov:2007ep} and their citations), while when $F(\phi)=\phi/M_{Pl}$ and $Z(\phi)=\omega_{BD}M_{Pl}/\phi$, it reduces to the Brans-Dicke theory \cite{Brans:1961sx}. The equation of motion for $\phi$ is: \be\label{eomJ} \ddot\phi+3H_J\dot\phi+\frac{Z_\phi}{2Z}\dot\phi^2-\frac{6M_{Pl}^2F_\phi}{Z}(\dot H+2H^2)+\frac{U_\phi}{Z}=0~,\ee where subscript ``$\phi$" indicates $\partial/\partial\phi$, and dot denotes derivative with respect to cosmic time in the Jordan frame, $t_J$. The ``total stress energy tensor" ${\cal E}_{\mu\nu}$ of the action (\ref{actionJ}) is defined as: \bea \delta_{g^{\mu\nu}}{\cal S}_J&=&\int d^4x\sqrt{-g}\delta g^{\mu\nu}{\cal E}_{\mu\nu}~,\nonumber\\ {\cal E}_{\mu\nu}&=&-\frac{1}{2}g_{\mu\nu}\Bigl[M_{Pl}^2FR-\frac{1}{2}Z(\nabla\phi)^2-U\Bigr]+M_{Pl}^2FR_{\mu\nu} \nonumber\\ &&-\frac{1}{2}Z\partial_\mu\phi\partial_\nu\phi+M_{Pl}^2F_{,\lambda;\varrho}(g^{\lambda\varrho}g_{\mu\nu}-\delta_\mu^\lambda\delta_\nu^\varrho)~. \eea By letting ${\cal E}_{\mu\nu}=0$, we can get the Friedmann Equation as: \be\label{friedmannJ} 6M_{Pl}^2H_J\dot F+6M_{Pl}^2H_J^2F=\frac{1}{2}Z\dot\phi^2+U~,\ee and by defining the energy density $\rho_J$ and the pressure $P_J$ to be \be \rho_J=3M_{Pl}^2H_J^2~,~P_J=-M_{Pl}^2(2\dot H_J+3H_J^2)~,\ee we can express $\rho_J$ and $P_J$ as: \bea\label{rhoJ} \rho_J&=&\frac{1}{2F}\big(\frac{1}{2}Z\dot\phi^2+U-6M_{Pl}^2H_J\dot F\big)~,\\ P_J&=&\frac{1}{2F}\big[\frac{1}{2}Z\dot\phi^2-U+2M_{Pl}^2(\ddot F+2H_J\dot F)\big]~.\eea

Making use of Eq. (\ref{conformal}) where $\Omega^2\equiv 2F$, We can transform the action (\ref{actionJ}) into its Einstein frame presentation. The corresponding action is: \be\label{actionE} {\cal S}_E=\int d^4\h{x}\sqrt{-\h{g}}\Bigl[\frac{M_{Pl}^2}{2}\h{R}-\frac{1}{2}(\h{\partial}\varphi_E)^2-V(\varphi_E)\Bigr]~,\ee which, as we mentioned, can act as that of a minimal coupling single scalar field. $\varphi_E$ is the redefined field in the Einstein frame, while $V(\varphi_E)$ is the potential of $\varphi_E$ in the Einstein frame. The transformation between actions (\ref{actionJ}) and (\ref{actionE}) is a straightforward and standard process, which could be found in the literatures, such as \cite{Faraoni:1998qx}. Here we only summarize the relations of some basic variables between the two frames as follows: \bea ~dt_E&=&\sqrt{2F}dt_J~,~a_E=\sqrt{2F}a_J~,~H_E=\frac{H_J}{\sqrt{2F}}(1+\frac{\dot F}{2H_JF})~,\nonumber\\\label{relation} \varphi_E&=&\int\sqrt{\frac{3M_{Pl}^2F_\phi^2+FZ}{2F^2}}d\phi~,~V(\varphi_E)=\frac{U(\phi)}{4F^2}~.\eea

Varying action (\ref{actionE}) with the field $\varphi_E$, we can get the equation of motion for $\varphi_E$, which is just that of a minimal coupling scalar field: \be\label{eomE} \frac{d^2\varphi_E}{dt_E^2}+3H_E\frac{d\varphi_E}{dt_E}+V_{\varphi_E}=0~,\ee and the energy density and pressure of $\varphi_E$ can also easily be obtained from action (\ref{actionE}), which are: \be\label{rhoandpE} \rho_E=\frac{1}{2}(\frac{d\varphi_E}{dt_E})^2+V~,~~~P_E=\frac{1}{2}(\frac{d\varphi_E}{dt_E})^2-V~.\ee Finally, the Friedmann Equations read: \be\label{friedmannE} 3M_{Pl}^2H_E^2=\rho_E~,~M_{Pl}^2\frac{dH_E}{dt_E}=-\frac{1}{2}(\rho_E+P_E)~.\ee

\subsection{perturbations}
As is well-known, cosmological perturbations are of great importance in the study of the early universe. It is believed that primordial quantum fluctuations, which is formed deep inside the horizon at the very beginning of the universe, could be stretched out of the horizon and become classical perturbations that provides the seedings of the current large scale structures in our universe. Theory of primordial perturbations has been well established from long time ago (see e.g. \cite{Kodama:1985bj,Mukhanov:1990me,Acquaviva:2002ud,Malik:2008im} for reviews), and observations have been imposed to put on constraints upon it \cite{Komatsu:2010fb,Larson:2010gs}. One of its most important signature is the scale-invariance of the 2-point correlation function of the curvature perturbation, i.e., power spectrum, which comes from the second-order perturbed action of the cosmological models. Now we turn on to the relations between the perturbation theories of a nonminimal coupling model in the two frames. In our case, we first perturb our action (\ref{actionJ}) in the Jordan frame to the quadratic-order in terms of the curvature perturbation ${\cal R}$. The curvature perturbation can be defined in the metric in the Arnowitt-Deser-Misner (ADM) form \cite{Arnowitt:1962hi}: \be \label{adm} ds^2=-{\cal N}^2(t,{\bf x})dt^2+2a^2(t){\cal N}_i(t,{\bf x})dtd{\bf x}^i+a^2(t)e^{2{\cal R}(t,{\bf x})}d{\bf x}^2~,\ee where ${\cal N}$ and ${\cal N}_i$ are the lapse function and shift vector respectively, and ${\bf x}$ denotes the spatial coordinates. Since under the conformal transformation (\ref{conformal}) the transformation factor $\Omega^2$ can be fully absorbed into the scale factor (see Eq. (\ref{relation})), the perturbation variables such as ${\cal R}$ is thus conformal invariant. The quadratic action can be written down as:  \be\label{pertactionJ} {\cal
S}^{C(2)}_J=\int d\eta d^3xa^2\frac{Q_{\cal R}}{c_{J}^2}\Bigl[{\cal R}^{\prime 2}-c_{J}^2(\partial{\cal R})^2\Bigr]~,\ee where we define \bea\label{QRJ} Q_{\cal R}&\equiv&\frac{2F}{(2+\delta_F)^2}[3\delta_{F}^{2}+\frac{\dot\phi^2Z}{H_J^{2}F}]~,\\ \label{csJ} c_{J}^2&\equiv&\frac{(\delta_{F}+2\epsilon_J)(2+\delta_{F})}{[3\delta_{F}^{2}+\frac{\dot\phi^2Z}{H_J^{2}F}]}
-\frac{2\dot{\delta}_{F}}{H_J[3\delta_{F}^{2}+\frac{\dot\phi^2Z}{H_J^{2}F}]}~,\eea with $\delta_F\equiv\dot F/H_JF$ and $\epsilon_J\equiv-\dot H_J/H_J^2$, and the prime denotes derivative with respect to the conformal time $\eta=\int {a_J}dt_J$. Varying (\ref{pertactionJ}) with respect to ${\cal R}$, one can straightforwardly write down the equation of motion for the perturbation as: \be\label{perteomCJ} \frac{d^2u_{{\cal R}J}}{dy^2}+(k^2-\frac{1}{z_{{\cal R}J}}\frac{d^2z_{{\cal R}J}}{dy^2})u_{{\cal R}J}=0~,\ee through the redefined variables $u_{{\cal R}J}=z_{{\cal R}J}{\cal R}$ and $z_{\cal R}\equiv a_J\sqrt{2Q_{\cal
R}/c_{J}}$, while \be y\equiv\int c_{J}d\eta~,\ee which is also conformal invariant according to Eq. (\ref{relation}).

The solution of Eq. (\ref{perteomCJ}) splits into subhorizon and superhorizon parts. The subhorizon solution of the above equation, where it is assumed that the $k^2$ term dominates over the last term in the bracket, is straightforward. By imposing the canonical quantization condition: \be u_{{\cal R}J}^{\ast}({\bf k})u_{{\cal R}J}^{\prime}({\bf k}^{\prime})-u_{{\cal R}J}({\bf k})u_{{\cal R}J}^{\ast\prime}({\bf k}^{\prime})=-i\delta^3({\bf k}-{\bf k}^\prime)~,\ee we obtain the following solution: \be u_{{\cal R}J}=\frac{1}{\sqrt{2c_Jk}}e^{iky}~,~{\cal R}=\frac{u_{{\cal R}J}}{z_{{\cal R}J}}=\frac{1}{a_J\sqrt{2Q_{\cal R}}}\frac{1}{\sqrt{2k}}e^{iky}~.\ee

The superhorizon solution is a little bit more complicated since it involves $z_{{\cal R}J}$, however, we can apply some ansatz to have it simplified. For example, we can assume that $z_{{\cal R}J}\sim |y|^{\lambda_C}$, then the superhorizon solution of Eq. (\ref{perteomCJ}) can be obtained as: \bea\label{resultCJ}
u_{{\cal R}J}&\sim&\sqrt{|y|}[c_1J_{{\lambda_C}-\frac{1}{2}}(k|y|)+c_2J_{\frac{1}{2}-{\lambda_C}}(k|y|)]\nonumber\\  &\sim& c_1k^{{\lambda_C}-\frac{1}{2}}|y|^{{\lambda_C}-\frac{1}{2}}+c_2k^{\frac{1}{2}-{\lambda_C}}|y|^{1-{\lambda_C}}~,\nonumber\\
{\cal R}&=&\frac{u_{{\cal R}J}}{z_{{\cal R}J}}\sim c_1k^{{\lambda_C}-\frac{1}{2}}+c_2k^{\frac{1}{2}-{\lambda_C}}|y|^{1-2{\lambda_C}}~,\eea where $J_i$ is the Bessel function and $c_1$, $c_2$ are constants.

Power spectrum is defined as \be {\cal P}^J_{\cal R}(k)\equiv\frac{k^3}{2\pi^2}\big|{\cal R}\big|^2=\frac{k^3}{2\pi^2}\bigg|\frac{u_{{\cal R}J}}{z_{{\cal R}J}}\bigg|^2~.\ee As one can see from (\ref{resultCJ}), scale-invariant spectrum can be obtained when ${\lambda_C}=2$ or $-1$, depending on whether the constant mode or the time-varying mode will be dominant, that is, on whether $|y|^{1-2{\lambda_C}}$ is decreasing or increasing with respect to the cosmic time $t_J$. In the usual case where one can ignore the variation of $c_{J}$ w.r.t. $t_J$, one therefore gets $y\simeq c_{J}\int a_J^{-1}dt_J\equiv c_{J}(\eta_\ast-\eta)$, then $z_{{\cal R}J}\sim|\eta_\ast-\eta|^{-1}$ is for the universe where the time-varying mode becomes decaying while the constant mode dominates the perturbation, such as inflation, while $z_{{\cal R}J}\sim|\eta_\ast-\eta|^2$ is for the universe where the time-varying mode is the growing mode and thus dominates over the constant one, such as matter-contraction. See also \cite{ArmendarizPicon:2003ht,Piao:2006ja,Magueijo:2008pm,Khoury:2008wj,Bessada:2009ns,Cai:2009hc} for the discussions on the cases with varying sound speed.

Other than the curvature perturbation, the tensor perturbation is another important perturbation which comes from the tensor component of the perturbed metric, associated with the production of the gravitational waves in the early universe. Similar to the curvature perturbation, we can also write down the second-order action for the  tensor perturbation as: \be {\cal S}^{T(2)}_J=\frac{1}{4}\int d\eta d^3xa_J^2F[h^{\prime2}-(\partial h)^2]~,\ee where $h$ is the polarization component of the tensor part of the perturbed metric $h_{ij}$ which is defined as: \be \label{admten} ds^2=-dt^2+a^2(t)e^{h_{ij}(t,{\bf x})}d{\bf x}^id{\bf x}^j~.\ee Note that since the sound speed of tensor perturbation is unity, there is no need to redefine variables such like $y$ for the curvature perturbation. For the same reason of the curvature perturbation, the tensor perturbation is also conformal invariant. The equation of motion for tensor perturbation can also be very easily written as: \be\label{perteomTJ} v_J^{\prime\prime}+(k^2-\frac{(a_J\sqrt{2F})^{\prime\prime}}{a_J\sqrt{2F}})v_J=0~,\ee where $v_J\equiv z_{TJ}h$ is a redefined variable through $z_{TJ}=a_J\sqrt{2F}$.

Similar to the above analysis, the subhorizon solution of Eq. (\ref{perteomTJ}) where $k^2$ dominates over the last term in the bracket is:  \be v_J=\frac{1}{\sqrt{2k}}e^{ik\eta}~,~h=\frac{v_J}{z_{TJ}}=\frac{1}{a_J\sqrt{2F}}\frac{1}{\sqrt{2k}}e^{ik\eta}~,\ee where we have imposed the canonical quantization condition $v_J^{\ast}({\bf k})v_J^{\prime}({\bf k}^{\prime})-v_J({\bf k})v_J^{\ast\prime}({\bf k}^{\prime})=-i\delta^3({\bf k}-{\bf k}^\prime)$. The superhorizon solution of Eq. (\ref{perteomTJ}) is: \bea\label{resultTJ} v_J &\sim& g_1k^{{\lambda_T}-\frac{1}{2}}|\eta_\ast-\eta|^{{\lambda_T}-\frac{1}{2}}+g_2k^{\frac{1}{2}-{\lambda_T}}|\eta_\ast-\eta|^{1-{\lambda_T}}~,\nonumber\\ h&=&\frac{v_J}{z_{TJ}}\sim g_1k^{{\lambda_T}-\frac{1}{2}}+g_2k^{\frac{1}{2}-{\lambda_T}}|\eta_\ast-\eta|^{1-2{\lambda_T}}~,\eea where we assumed $z_{TJ}\sim |\eta_\ast-\eta|^{\lambda_T}$ and $g_1$ and $g_2$ are constants. Moreover, the power spectrum of the tensor perturbation is also defined as: \be {\cal P}^J_{T}(k)\equiv\frac{k^3}{2\pi^2}|h|^2=\frac{k^3}{2\pi^2}\bigg|\frac{v_J}{z_{TJ}}\bigg|^2~,\ee therefore one can see that, the scale-invariance of the tensor perturbation requires $a_J\sqrt{2F}\sim |\eta_\ast-\eta|^{-1}$ for the case where the constant mode dominates, or $a_J\sqrt{2F}\sim |\eta_\ast-\eta|^2$ for the case where increasing mode dominates.

Combining the two requirements for the scale-invariant spectrum of the curvature and tensor perturbations, we can conclude that in order to satisfy both of them, we must require that {\it i)} $a_J\sqrt{2F}\sim |\eta_\ast-\eta|^{-1}$ or $a_J\sqrt{2F}\sim |\eta_\ast-\eta|^2$, and {\it ii)} $\sqrt{Q_{\cal R}/c_{J}F}$ is nearly constant. As we will see below, for the first condition, the first case corresponds to de Sitter expansion (inflation) while the second case corresponds to the matter contraction in the Einstein frame. For the second condition, if in addition we assume that the universe evolves with a constant sound speed $c_{J}$, which is used above to relate $y$ and $\eta$, then we also have to expect that $Q_{\cal R}$ is proportional to $F$. Actually, as can be seen in the next section, both these requirements can be very easily satisfied.

The above analysis is for the perturbations of a nonminimal coupling theory in the Jordan frame. In the Einstein frame, however, the action of the theory follows Eq. (\ref{actionE}), which is very much like that of a minimal coupling field in GR. Following the same steps, we can obtain the second-order perturbed action of Eq. (\ref{actionE}) in terms of the curvature perturbation as: \be\label{pertactionE} {\cal S}^{C(2)}_E=\int d\eta d^3xa_E^2\epsilon_E[{\cal R}^{\prime 2}-(\partial{\cal R})^2]~,\ee where $\epsilon_E$ is defined as \be\label{epsilonE} \epsilon_E\equiv-H_E^{-2}\frac{dH_E}{dt_E}=\frac{3}{2}(1+w_E)\ee with $w_E$ being the equation of state of the universe in the Einstein frame. Note also that the sound speed squared in the Einstein frame, namely $c_{E}$, turns out to be unity. Actually, if one directly transforms the quantities $Q_{\cal R}$ and $c_{J}^2$ in Eqs. (\ref{QRJ}) and (\ref{csJ}) into the Einstein frame, the action (\ref{pertactionE}) can also be obtained, namely we have: \be\label{relation2} a_J^2Q_{\cal R}\rightarrow a_E^2\epsilon_E~,~c_{J}^2\rightarrow 1~\ee by doing conformal transformation. This shows that the two frames are equivalent in perturbation level, and could lead to the same perturbative behavior. According to Eq. (\ref{pertactionE}), one can straightforwardly write down the equation of motion of perturbation as: \be u^{\prime\prime}_{{\cal R}E}+(k^2-\frac{z^{\prime\prime}_{{\cal R}E}}{z_{{\cal R}E}})u_{{\cal R}E}=0~,\ee where $u_{{\cal R}E}=z_{{\cal R}E}{\cal R}$ and $z_{{\cal R}E}\equiv a_E\sqrt{2\epsilon_E}$. Even without doing further calculations, one can learn from the above analysis that scale-invariant spectrum can be obtained when $z_{{\cal R}E}\sim|\eta_\ast-\eta|^{-1}$ or $z_{{\cal R}E}\sim|\eta_\ast-\eta|^2$, depending on whether the constant mode or the time-varying mode will dominate over, or in other words, on whether the time-varying mode is decreasing or increasing with respect to cosmic time $t_E$.

We can also perform the same analysis to the tensor perturbation in the Einstein frame. The second-order action can be written down as: \be\label{pertenactionE} {\cal S}^{T(2)}_E=\frac{1}{4}\int d\eta d^3xa_E^2[h^{\prime2}-(\partial h)^2]~,\ee while the equation of motion for tensor perturbation is: \be v_E^{\prime\prime}+(k^2-\frac{a_E^{\prime\prime}}{a_E})v_E=0~,\ee where we defined $v_E\equiv z_{TE}h$ and $z_{TE}=a_E=\sqrt{2F}a_J$. Again, the scale-invariance of tensor perturbation requires $a_E\sim |\eta_\ast-\eta|^{-1}$ for the case where the constant mode dominates, while $a_E\sim |\eta_\ast-\eta|^2$ for the case where the increasing mode dominates.

To summarize, one can see that in order to have both curvature and tensor perturbations scale-invariant in the Einstein frame, we must require that {\it i)} $a_E\sim |\eta_\ast-\eta|^{-1}$ or $a_E\sim |\eta_\ast-\eta|^2$, and {\it ii)} $\epsilon_E$ is nearly constant. The two conditions can be directly derived from the two conditions in the Jordan frame, which is due to the equivalence of the two frames, however here they can be more clearly connected to the evolution behavior of the universe. For example, one can see easily that for the first condition, the first case refers to an expanding universe with $w_E\simeq-1$, which is nothing but inflation, while the second case corresponds to the universe contracting with $w_E\simeq0$, which is nearly dust-like, or in a ``matter-contracting" phase. Note that the latter one, followed by a nonsingular transfer to a normal expanding universe, has been widely applied in bouncing cosmology, which is expected to act as an alternatives of inflation cosmology \cite{Cai:2007qw,Novello:2008ra}. The second condition can also be easily satisfied, if we set $w_E$ to be nearly constant, which is a very common and reasonable assumption. Note that for more general cases with its sound speed $c_{E}\neq1$, however, it should be required that $\sqrt{\epsilon_E/c_{E}}$ be nearly constant, rather than $\epsilon_E$ only.

\section{Reconstruction of a Nonminimal Coupling Theory with Scale-invariant Power Spectrum}
From the last section, we know that Jordan and Einstein frames are equivalent at the perturbation level, therefore the conditions for a nonminimal coupling theory in its counterpart Jordan frame to have scale-invariant power spectrum can be mapped into the conditions in the Einstein frame, where the theory can be viewed as that of a minimal coupling field in GR. Since the solution for a minimal coupling field with scale-invariant power spectrum are well-known (for example, an inflationary solution or a matter-contracting solution), one could reconstruct the solutions for the nonminimal coupling theory in its counterpart Jordan frame, and obtain what forms of the action of the field can naturally lead to these solutions. This is our main goal of this section. We will reconstruct the nonminimal coupling theory from both inflationary and matter-contracting scenarios separately, and obtain the form of the actions as well as the dependence of the universe evolution (i.e., the equation of state) on the action parameters.
\subsection{Reconstruction from inflation}\label{recinf}
\subsubsection{A minimal coupling inflation model}
First of all we consider the inflationary solution generated from a minimal coupling single scalar field model, the action of which is given by Eq. (\ref{actionE}). The inflationary solution possesses the property of having its equation of state $w_E\simeq-1$, thus according to the Friedmann Equations (\ref{friedmannE}) one can parameterize the scale factor $a_E$ and the Hubble parameter $H_E$ in terms of $t_E$ as: \be \label{aandHI} a_E(t_E)=a_\ast\Big(\frac{t_E}{t_E^\ast}\Big)^{1/\epsilon_E}~,~H_E=\frac{1}{\epsilon_Et_E}~,\ee where the slow-roll parameter $\epsilon_E$ is defined in Eq. (\ref{epsilonE}), and for the inflation case one generally have $0<\epsilon_E\ll1$ so that $a_E$ have a large power-law index of $t_E$ \footnote{Inflation could also be phantom-like \cite{Piao:2003ty,Piao:2004tq,Lidsey:2004xd,GonzalezDiaz:2004df,Baldi:2005gk,Nojiri:2005pu,Wu:2006wu,Piao:2007ne,Liu:2012ib}, in which case one has $\epsilon_E<0$, and the parameterizations of $a_E$ and $H_E$ is also slightly modified from Eq. (\ref{aandHI}).} . Note that since the universe are expanding, $t_E$ goes from $0_+$ to $+\infty$, and we set $a=a_\ast$ at the time point when $t_E=t_E^\ast$. Substitute this parametrization into Eqs. (\ref{rhoandpE}) and (\ref{friedmannE}) we can have a natural and self-consistent solution of $\varphi_E$: \be\label{varphiEI}
\varphi_E=\sqrt{\frac{2}{\epsilon_E}}M_{Pl}\ln(M^{-1}t_E)-\sqrt{\frac{2}{\epsilon_E}}M_{Pl}\ln(M^{-1}t^\ast_E)+\varphi^\ast_E~,\ee where $M$ is some energy scale, while $\varphi^\ast_E$ and $t^\ast_E$ are integral constants, and can be easily eliminated by setting $\varphi^\ast_E=\sqrt{2/\epsilon_E}M_{Pl}\ln(M^{-1}t^\ast_E)$. The potential could be arbitrary, however, here for sake of simplicity, we just take the potential that can give rise to a scaling solution. From the definition of the equation of state of $\varphi_E$: \be w_E=\frac{(d\varphi_E/dt_E)^2-2V}{(d\varphi_E/dt_E)^2-2V}=-1+\frac{2}{3}\epsilon_E~,\ee we can get the form of the potential: \be V(\varphi_E)\sim \frac{(3-\epsilon_E)}{\epsilon_E^2}\frac{M_{Pl}^2}{t_E^2}\sim\frac{(3-\epsilon_E)}{\epsilon_E^2}M^2M_{Pl}^2e^{-\sqrt{2\epsilon_E}\frac{\varphi_E}{M_{Pl}}}~.\ee

We could write the potential generally as \be \label{potentialEI} V(\varphi_E)=V_0e^{-\sqrt{2\epsilon_E}\varphi_E/M_{Pl}}~,\ee where $V_0$ is some constant with dimension of 4. In general, to have scaling solution, an exponential potential is required.

\subsubsection{Its nonminimal coupling correspondence}
Now we focus on finding the proper form of action (\ref{actionJ}) that can give rise to scale-invariant power spectrum, by reconstructing it from the above inflation model. To do this, we need to assume that the above model is the presentation of the required model in its Einstein frame. First of all, we parameterize the function $F(\phi)$ in Eq. (\ref{actionJ}) in terms of $t_J$: \be\label{parametrizeFI} F(\phi(t_J))=F_\ast\Big(\frac{|t_J|}{|t^\ast_J|}\Big)^{2f_I}~,\ee where $f_I$ is some parameter. From Eq. (\ref{relation}) it follows that \be\label{relationtI}
\int^{t^\ast_E}_{t_E}dt_E=\int^{t^\ast_J}_{t_J}\sqrt{2F}dt_J=\frac{\sqrt{2F_\ast}}{|t^\ast_J|^{f_I}}\int^{t^\ast_J}_{t_J}|t_J|^{f_I}dt_J~.\ee Note that here we assume that the same as $t_E$, $t_J$ monotonically increases, although its value can be either positive or negative. This is an arbitrary choice, only indicating the arrow of time, and one can surely assume that time goes in an opposite direction, which is only trivially dual to the current case by the transformation $t_J^\prime\rightarrow-t_J$. Considering this, there will be two different cases: one is $t_J>0$ while the other is $t_J<0$. In the $t_J>0$ case, we set $F(\phi)$ as \be F(\phi)=F_\ast\Big(\frac{t_J}{t^\ast_J}\Big)^{2f_I}~,\ee and we can obtain the relation between $t_E$ and $t_J$ making use of Eq. (\ref{relationtI}). Note that the relation is quite different for $f_I\neq-1$ and $f_I=-1$. For $f_I\neq-1$, we obtain \be t_E=\frac{\sqrt{2F_\ast}t^\ast_J}{f_I+1}\Big(\frac{t_J}{t^\ast_J}\Big)^{f_I+1}-\frac{\sqrt{2F_\ast}t^\ast_J}{f_I+1}+t^\ast_E~.\ee Since $t_E$ and $t_J$ are all increasing as assumed, in order to make the above equation consistent, $f_I>-1$ is required, and we can set $t^\ast_E=\sqrt{2F_\ast}t^\ast_J/(f_I+1)$ to get rid of the nonessential integral constants. For $f_I=-1$, we have $F=F_\ast(t_J/t^\ast_J)^{-2}$ and \be t_E=\sqrt{2F_\ast}t^\ast_J\ln\Big(\frac{t_J}{t_{Pl}}\Big)-\sqrt{2F_\ast}t^\ast_J\ln\Big(\frac{t^\ast_J}{t_{Pl}}\Big)+t^\ast_E~.\ee Here we use the Planck time $t_{Pl}$ as a dimensional renormalization in the logarithm function and we also set $t^\ast_E=\sqrt{2F_\ast}t^\ast_J\ln(t^\ast_J/t_{Pl})$. Note that unless $t^\ast_J<t_{Pl}$ which is very close to 0, $\ln(t^\ast_J/t_{Pl})$ will be positive which is consistent with this setting.

In the $t_J<0$ case, we have \be F(\phi)=F_\ast\Big(\frac{-t_J}{-t^\ast_J}\Big)^{2f_I}~,\ee and with Eq. (\ref{relationtI}) we have \be t_E=-\frac{\sqrt{2F_\ast}(-t^\ast_J)}{f_I+1}\Big(\frac{-t_J}{-t^\ast_J}\Big)^{f_I+1}+\frac{\sqrt{2F_\ast}(-t^\ast_J)}{f_I+1}+t^\ast_E~\ee with $t^\ast_E=-\sqrt{2F_\ast}(-t^\ast_J)/(f_I+1)$ for $f_I\neq-1$ and \be
t_E=-\sqrt{2F_\ast}(-t^\ast_J)\ln\Big(\frac{-t_J}{t_{Pl}}\Big)+\sqrt{2F_\ast}(-t^\ast_J)\ln\Big(\frac{-t^\ast_J}{t_{Pl}}\Big)+t^\ast_E~\ee with $t^\ast_E=-\sqrt{2F_\ast}(-t^\ast_J)\ln[(-t^\ast_J)/t_{Pl}]$ for $f_I=-1$. For $f_I\neq1$ case, in order to make the above equation consistent, $f_I<-1$ is required. However, for $f_I=-1$ case, as is opposite to that for $t_J>0$, unless $-t^\ast_J<t_{Pl}$ which is meaningless, $\ln[(-t^\ast_J)/t_{Pl}]$ is positive and make the setting for $t^\ast_E$ inconsistent. So we will not consider this case.

In summary, we have: \bea\label{tE2tJI} t_E=\left\{ \begin{array}{l} \frac{\sqrt{2F_\ast}t^\ast_J}{f_I+1}\Big(\frac{t_J}{t^\ast_J}\Big)^{f_I+1}~~~~{\rm for}~~f_I>-1~,\\\\ \sqrt{2F_\ast}t^\ast_J\ln\bar{t}_J~~~~{\rm for}~~f_I=-1~,\\\\ -\frac{\sqrt{2F_\ast}(-t^\ast_J)}{f_I+1}\Big(\frac{-t_J}{-t^\ast_J}\Big)^{f_I+1}~~~~{\rm for}~~f_I<-1~,\\ \end{array}\right. \eea where we also defined the dimensionless variable $\bar{t}_J\equiv t_J/t_{Pl}$.

Now we proceed to check the dependence of $a_J$, $H_J$ and $\phi$ on $t_J$. To this aim we will make use of Eq. (\ref{relation}). First of all, the scale factor in the Jordan frame $a_J$ is: \be a_J(t_J)=\frac{a_E}{\sqrt{2F}}=\frac{a^\ast_E}{{t^\ast_E}^{1/\epsilon_E}}\frac{t_E^{1/\epsilon_E}}{\sqrt{2F}}~,\ee and to describe this in terms of $t_J$, we need to use Eqs. (\ref{parametrizeFI}) and (\ref{tE2tJI}) according to different cases. For $f_I>-1$, we have: \bea\label{a2tJfgt-1I} a_J(t_J)&=&\frac{a^\ast_E{t^\ast_J}^{f_I}}{{t^\ast_E}^{1/\epsilon_E}\sqrt{2F_\ast}}\frac{t_E^{1/\epsilon_E}}{t_J^{f_I}}~,\nonumber\\
&=&\frac{a^\ast_E(2F_\ast)^{\frac{1-\epsilon_E}{2\epsilon_E}}}{(f_I+1)^{1/\epsilon_E}}\Big(\frac{t^\ast_J}{t^\ast_E}\Big)^{\frac{1}{\epsilon_E}}\Big(\frac{t_J}{t^\ast_J}\Big)^{\frac{f_I(1-\epsilon_E)+1}{\epsilon_E}}~.\eea Similarly, we have:\be\label{a2tJfeq-1I} a_J(t_J)=a^\ast_E(2F_\ast)^{\frac{1-\epsilon_E}{2\epsilon_E}}\Big(\frac{t^\ast_J}{t^\ast_E}\Big)^{\frac{1}{\epsilon_E}}\Big(\frac{t_J}{t^\ast_J}\Big)(\ln\bar{t}_J)^\frac{1}{\epsilon_E}~,\ee for $f_I=-1$, and \be\label{a2tJflt-1I} a_J(t_J)=\frac{a^\ast_E(2F_\ast)^{\frac{1-\epsilon_E}{2\epsilon_E}}}{(f_I+1)^{1/\epsilon_E}}\Big(\frac{-t^\ast_J}{t^\ast_E}\Big)^{\frac{1}{\epsilon_E}}\Big(\frac{-t_J}{-t^\ast_J}\Big)^{\frac{f_I(1-\epsilon_E)+1}{\epsilon_E}}~,\ee for $f_I<-1$, respectively. In deriving all these formulae, we have applied Eq. (\ref{tE2tJI}) for $t_E$.

\begin{figure}[htbp]
\centering
\includegraphics[scale=0.3]{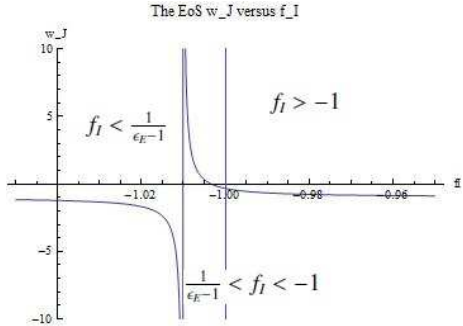}
\caption{The behavior of $w_J$ in $f_I\neq-1$ case w.r.t. $t_J$. The whole space is divided into 3 parts, namely where $f_I>-1$ or $f_I<1/(\epsilon_E-1)$ the universe expands while where $1/(\epsilon_E-1)<f_I<-1$ it contracts. Here we set the value of $\epsilon_E$ to be $\epsilon_E=0.01$ as an example. The points of $f_I=1/(\epsilon_E-1)$ corresponds to $w_J\rightarrow\pm\infty$ while $f_I=-1$ corresponds to $w_J=-1/3$. When $f_I=0$, $w_J$ is equal to $-1+2\epsilon_E/3$ in the expanding region, which is trivial. The approach of $f_I$ to $1/(\epsilon_E-1)$ from left and right hand side leads $w_J$ to negative and positive infinity, which corresponds to slow contraction or expansion, while $f_I\rightarrow\infty$ leads $w_J$ to $-1$ in expanding region, which is inflation.}\label{wJvsfI}
\end{figure}

Following the same approach, we can also determine the $t_J$-dependence of $H_J$ and $\phi$, as well as the equation of state $w_J$, since $w_J$ can be related to $H_J$ through the relationship $w_J=-1-2\dot H_J/3H_J^2$. From Eq. (\ref{relation}), one gets: \be\label{relationH} H_J=(H_E-\frac{\dot F}{2F^{3/2}})\sqrt{2F}=\frac{2}{3t_E}\sqrt{2F}-\frac{\dot F}{2F}~,\ee and substitute Eqs. (\ref{parametrizeFI}) and (\ref{tE2tJI}) into the above formula, we get: \bea \label{H2tJI} H_J=\left\{ \begin{array}{l} \frac{f_I(1-\epsilon_E)+1}{\epsilon_E t_J}~~~~{\rm for}~~f_I>-1~{\rm and}~f_I<-1~,\\\\ \frac{1}{t_J}\big(\frac{1}{\epsilon_E\ln\bar{t}_J}+1\big)~~~~{\rm for}~~f_I=-1~.\\ \end{array}\right.\eea In the first case, $\epsilon_J$ can also be written as $\epsilon_J\equiv\epsilon_E/[f_I(1-\epsilon_E)+1]$, so as to have $H_J=1/(\epsilon_Jt_J)$. From the above one can see that, besides $f_I=-1$, another important point which divides different behaviors of the universe is to have $f_I(1-\epsilon_E)+1=0$, namely $f_I=1/(\epsilon_E-1)$. For normal inflation with $\epsilon_E>0$, $1/(\epsilon_E-1)<-1$, this point is on the left hand side of that for $f_I=-1$. With this two dividing points, we can separate the whole parameter space of the universe evolution in the Jordan frame into three parts. For example, when $f_I>-1$, we have $f_I(1-\epsilon_E)+1>0$ and $t_J>0$, then the Hubble parameter $H_J>0$, indicating also an expanding universe in the Jordan frame. When $f_I<1/(\epsilon_E-1)$, we have both $f_I(1-\epsilon_E)+1<0$ and $t_J<0$, so $H_J$ is still positive, and the universe in the Jordan frame is also expanding. However, when $1/(\epsilon_E-1)<f_I<-1$, we will get $f_I(1-\epsilon_E)+1>0$ and $t_J<0$, leading to a contracting universe in the Jordan frame. That means with nonminimal coupling, one can even get a contracting evolution of the universe, which corresponds to inflation in its Einstein frame, but since $\epsilon_E\ll 1$, this parameter space will be very narrow, leading to a small probability. Finally, we can also see from Eq. (\ref{H2tJI}) that in the case $f_I=-1$, the universe is also expanding due to the positive $H_J$. The same conclusions can also be obtained from the viewpoint of $a_J$: one can see from Eqs. (\ref{a2tJfgt-1I}-\ref{a2tJflt-1I}) that $a_J(t_J)$ is expanding when $f_I\geq-1$ or $f_I<1/(\epsilon_E-1)$, while contracting when $1/(\epsilon_E-1)<f_I<-1$. The equation of state in each case can also be derived as: \bea \label{w2tJI} w_J=\left\{ \begin{array}{l} \frac{3f_I(\epsilon_E-1)+2\epsilon_E-3}{3[f_I(1-\epsilon_E)+1]}~~~~{\rm for}~~f_I>-1~{\rm and}~f_I<-1~,\\\\ -\frac{\epsilon_E^2(\ln\bar{t}_J)^2+4\epsilon_E\ln \bar{t}_J-2\epsilon_E+3}{3(1+\epsilon_E\ln\bar{t}_J)^2}~~~~{\rm for}~~f_I=-1~,\\ \end{array}\right.\eea and the relation between $w_J$ and $f_I$, as well as the region where the universe contracts or expands is shown in Fig. \ref{wJvsfI}.

\begin{figure}[htbp]
\centering
\includegraphics[scale=0.5]{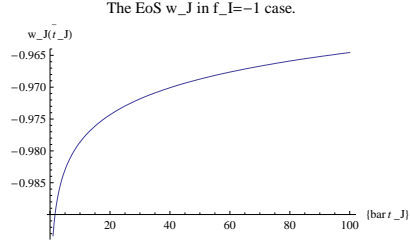}
\caption{The behavior of $w_J$ in $f_I=-1$ case w.r.t. $\bar{t}_J$. In this case, $w_J$ is no longer a constant, especially near $\bar{t}_J=0$, but after long time expansion when $\bar{t}_J\gg1$, it will approach to a constant value. The value of $\epsilon_E$ has been chosen to be $\epsilon_E=0.01$.}\label{wJIfeq-1}
\end{figure}

From the above results we can see that, due to the conformal equivalence of Jordan and Einstein frames, if the nonminimal coupling theory with the nonminimal factor $F(\phi)$ parameterized as Eq. (\ref{parametrizeFI}), while the evolution of the universe and the parameter $f_I$ satisfy the relation like Eqs. (\ref{a2tJfgt-1I}-\ref{a2tJflt-1I}), (\ref{H2tJI}) or (\ref{w2tJI}) with some $\epsilon_E$, the perturbations generated from such a theory should be scale-invariant. For the case where $f_I\neq-1$, it is quite obvious: since the scale factor of the universe behaves as a power-law function of $t_J$, which leads to a constant equation of state $w_J$, the slow-roll parameter in the Jordan frame $\epsilon_J$ is also a constant. Moreover, since $\delta_F$ is also constant, which is due to the power-law scaling of $F(\phi)$, we can easily make our $c_{J}$ to be constant, thus the condition that $\sqrt{Q_{\cal R}/c_{J}F}$ being constant can be easily satisfied making use of Eqs. (\ref{QRJ}) and (\ref{csJ}), as long as we choose proper form of $Z(\phi)$. Note that another condition, namely $a_J\sqrt{2F}\sim |\eta_\ast-\eta|^{-1}$, has already been satisfied by the inflation background in the Einstein frame with the conformal relation (\ref{relation}). However, it may not be so clear for the $f_I=-1$ case, since in this case, $w_J$ is no longer a constant, nor are $\epsilon_J$ and $c_J$. Nonetheless, we can see from Fig. \ref{wJIfeq-1} that for $\bar{t}_J\gg1$, that is, when the expansion has lasted for sufficiently long time, $w_J$ will approach to a constant value. That is, if for the perturbations created in sufficient late time, rough constant $\epsilon_J$ could still be obtained which can obviously satisfy our scale-invariance conditions.

As a side remark, we should mention that people having first glance at Fig. \ref{wJvsfI} may worry that due to the difference of the equation of state between the two frames, $w_J$ might be too large (e.g. $w_J\geq 1$) such that the universe in Jordan frame would suffer from some conceptual problems like inhomogeneities etc., while in Einstein frame not. This is actually not true, however. The reason is, though it seems that $w_J$ differs so much from $w_E$, the difference cannot be arbitrary. For example, for nearly de-Sitter-like $w_E$, from Fig. 1 we can see that the corresponding $w_J$ will never exceed the bound $w_J=-1/3$ when it describes an expanding universe in Jordan frame, while also never go below that bound when it stands for a contracting one. It is well known that in both case the inhomogeneity problem will not happen. This is a very important observation, which also supports our motivation that the two frames are equivalent and correspondable. The same fact also holds for the matter-contraction case. It will also be emphasized in the next section.

Finally let's discuss the dependence of $\phi$ on $t_J$, and as we will see later, by doing this we can not only determine $f_I$ and furthermore the equation of state $w_J$, but also get the form of the function $F(\phi)$ and the potential $U(\phi)$ in terms of $\phi$, thus the counterpart action (\ref{actionJ}) can at last be presented totally using the field variables, as it ought to be, instead of just parameterized forms. First of all, from Eq. (\ref{relation}) we know that: \bea\Big(\frac{d\varphi_E}{dt_E}\Big)^2&=&\frac{1}{2F}\Big(\frac{3M_{Pl}^2F_\phi^2+FZ}{2F^2}\Big)\Big(\frac{d\phi}{dt_J}\Big)^2~,\nonumber\\ &=&\frac{3M_{Pl}^2}{4F^3}\Big(\frac{dF}{dt_J}\Big)^2+\frac{Z\dot\phi^2}{4F^2}~,\nonumber\\ \label{dotphiE2J} &=&\frac{3M_{Pl}^2f_I^2}{(\pm t_J)^2F}+\frac{Z\dot\phi^2}{4F^2}~,\eea where we have made use of Eq. (\ref{parametrizeFI}), and the plus and minus sign in front of $t_J$ denotes the $f_I>-1$ and $f_I<-1$ cases, respectively. From Eq. (\ref{varphiEI}) we know that the left hand side of the above equation is \be (\frac{d\varphi_E}{dt_E})^2=\frac{2M_{Pl}^2}{\epsilon_Et_E^2}~,\ee then using Eqs. (\ref{tE2tJI}) we have: \bea \label{zdotphi2I} Z\dot\phi^2=\left\{ \begin{array}{l} \frac{4M_{Pl}^2F}{t_J^2}\big[\frac{(f_I+1)^2}{\epsilon_E}-3f_I^2\big]~~~~{\rm for}~~f_I>-1~,\\\\ \frac{4M_{Pl}^2F}{t_J^2}\big[\frac{1}{\epsilon_E(\ln\bar{t}_J)^2}-3\big]~~~~{\rm for}~~f_I=-1~,\\\\ \frac{4M_{Pl}^2F}{(-t_J)^2}\big[\frac{(f_I+1)^2}{\epsilon_E}-3f_I^2\big]~~~~{\rm for}~~f_I<-1~.\\ \end{array}\right. \eea

We assume that $Z(\phi)=Z_0^I\phi^{2z_I}$, then the left hand side of Eq. (\ref{zdotphi2I}) becomes \be Z\dot\phi^2\sim (\sqrt{Z_0^I}\phi^{z_I}\dot\phi)^2\sim \Big(\frac{\sqrt{Z_0^I}}{(z_I+1)}\frac{d\phi^{z_I+1}}{dt_J}\Big)^2~,\ee substituting it into Eq. (\ref{zdotphi2I}) one can get that for $f_I\neq-1$: \bea\label{phiJI} \phi^{z_I+1}&=&\frac{(z_I+1)}{f_I}\sqrt{\frac{4M_{Pl}^2F_\ast}{Z_0^I}\Big[\frac{(f_I+1)^2}{\epsilon_E}-3f_I^2\Big]}\Big(\frac{\pm t_J}{\pm t_J^\ast}\Big)^{f_I}~\nonumber\\ &=&\frac{(z_I+1)}{f_I}\sqrt{\frac{4M_{Pl}^2}{Z_0^I}\Big[\frac{(f_I+1)^2}{\epsilon_E}-3f_I^2\Big]}\sqrt{F}~.\eea Then we can at last present $F(\phi)$ in terms of $\phi$, which is: \bea \label{parametrizeF2I} F&=&F_0^I\phi^{2z_I+2}~,\\ F_0^I&=&\frac{Z_0^If_I^2}{4M_{Pl}^2(z_I+1)^2}\Big[\frac{(f_I+1)^2}{\epsilon_E}-3f_I^2\Big]^{-1}~.\eea

Note that from Eqs. (\ref{zdotphi2I}) and (\ref{parametrizeF2I}), one can also check that $Z\dot\phi^2/(H_J^{2}F)\sim\dot\phi^2/(H_J^{2}\phi^2)\sim {\rm const.}$, since $\phi$ is also a power-law function of $t_J$. Then from Eqs. (\ref{QRJ}) and (\ref{csJ}) we see that indeed $c_J$ is constant while $Q_{\cal R}\sim F$, as expected, so the above analysis on $\phi$ is totally consistent with the previous analysis on $a_J$, $H_J$ and $w_J$.

As the last step, we will reconstruct the form of the potential $U(\phi)$ in terms of $\phi$. To do this we have to make use of the equation of motion of $\phi$. First of all, from Eq. (\ref{phiJI}) we can express $\phi$ as: \bea \label{parametrizephiJI} \phi&=&\phi_0^I(\pm t_J)^{\frac{f_I}{z_I+1}}~,\\ \phi_0^I&=&\left\{\frac{(z_I+1)}{f_I(\pm t_J^\ast)^{f_I}}\sqrt{\frac{4M_{Pl}^2 F_\ast}{Z_0^I}\Big[\frac{(f_I+1)^2}{\epsilon_E}-3f_I^2\Big]}\right\}^\frac{1}{z_I+1}~,\eea so that $(\pm t_J)=(\phi/\phi_0^I)^{(z_I+1)/f_I}$. From Eq. (\ref{parametrizephiJI}), we can get the first and second time derivatives of $\phi$: \bea \label{parametrizedotphiJI} \dot\phi&=&\pm\frac{f_I}{z_I+1}\phi_0^I(\pm t_J)^\frac{f_I-z_I-1}{z_I+1}~,\\ \ddot\phi&=&\frac{f_I(f_I-z_I-1)}{(z_I+1)^2}\phi_0^I(\pm t_J)^\frac{f_I-2z_I-2}{z_I+1}~.\eea Then substituting it into either Eq. (\ref{eomJ}) or Eq. (\ref{friedmannJ}) we can get: \bea\label{potentialJI} U(\phi)&=&2\Big(\frac{3}{\epsilon_E^2}-\frac{1}{\epsilon_E}\Big)(f_I+1)^2\frac{F}{(\pm t_J)^2}~\nonumber\\ &=&U_0^I\phi^{2(z_I+1)(1-\frac{1}{f_I})}~,\\ U_0^I&=&2\Big(\frac{3}{\epsilon_E^2}-\frac{1}{\epsilon_E}\Big)(f_I+1)^2F_0^I(\phi_0^I)^\frac{2(z_I+1)}{f_I}~.\eea Comparing with the potential in Einstein Frame (\ref{potentialEI}) we find that it is also consistent with the relation (\ref{relation}).

From Eq. (\ref{potentialJI}), we see that for a given $z_I$ and $f_I$, the power-law index of $U(\phi)$ with respect of $\phi$ can be determined. From the opposite side, with given power-law forms of $U(\phi)$ and $Z(\phi)$, one can determine $f_I$, and therefore $w_J$, thus the evolution behavior in the Jordan frame can be determined. Since usually we start with the action with fixed $F(\phi)$, $Z(\phi)$ and $U(\phi)$, the latter case is more interesting to us. For example, if we start with the action of form (\ref{actionJ}), with \be\label{FandZandUI} F(\phi)=F_0^I\phi^{2z_I+2}~,~~~Z(\phi)=Z_0^I\phi^{2z_I}~,~~~U(\phi)=U_0^I\phi^{q_I}~,\ee then from Eq. (\ref{potentialJI}) we have \be\label{qIandfI} q_I=2(z_I+1)(1-\frac{1}{f_I})~,~f_I=\frac{2(z_I+1)}{2(z_I+1)-q_I}~,\ee and furthermore, from Eq. (\ref{w2tJI}), we can easily get: \be\label{wJI} w_J=\frac{2(z_I+1)(5\epsilon_E-6)-q_I(2\epsilon_E-3)}{3[2(z_I+1)(2-\epsilon_E)-q_I]}~.\ee

Eqs. (\ref{FandZandUI}-\ref{wJI}) are our main result for this subsection, linking the parameters $z_I$ and $q_I$ from the action (\ref{actionJ}) in the Jordan frame, to the behavior of the universe that it may drive, which corresponds to an inflation in the Einstein frame and thus give rise to scale-invariant power spectrum. From this relation we can see that, given the requirement of generating scale-invariant power spectrum, the evolution in the Jordan frame has more degrees of freedom than that in the Einstein frame, and is more dependent on the form of the action. For example, when $z_I=-1/2$ and $q_I=4$, we have the Brans-Dicke-like action where $F(\phi)\sim\phi$, $Z(\phi)\sim\phi^{-1}$ and $U(\phi)\sim\phi^4$, and in this case we get $w_J=(\epsilon_E-2)/(\epsilon_E+2)\simeq-1$, which behaves like inflation. Another interesting example is $z_I=0$, $q_I=4-2\epsilon_E/(1-\epsilon_E)$, which gives $F(\phi)\sim\phi^2$, $U(\phi)\sim\phi^{4[1-\epsilon_E/2(1-\epsilon_E)]}$ and $Z(\phi)$ is a constant. This gives $w_J=(1-2/\epsilon_E)/3$, which goes to deep below $-1$ for extremely small but positive $\epsilon_E$. This indicates that the universe is in a ``slow expansion" phase \cite{Piao:2003ty,Piao:2007sv,Joyce:2011ta}, which has been studied earlier in \cite{Piao:2011bz} with consistent conclusions with ours. Actually in the case where $|\epsilon_E|\ll1$, the correspondence can also hold for a range of $\epsilon_E$, with $|\epsilon_J|\simeq1/\epsilon_E$, providing a dual relation of this two scenarios \cite{Piao:2011bz}. For the case of $q_I=4-2\epsilon_E(1-\epsilon_E)/(1-2\epsilon_E-\epsilon_E^2)$ with the same $z_I$, however, a slow contracting universe with its equation of state as $w_J=-(7-2/\epsilon_E)/3$, which goes to highly above $1$ for small $\epsilon_E$, is given, which is kind of ``Ekpyrotic" scenario \cite{Khoury:2001wf}. Also note that the duality of both slow contraction and slow expansion to inflation has been discussed in \cite{Khoury:2010gw} within a different context.

The same approach could be done for $f_I=-1$ case. However, as can be seen from Eq. (\ref{tE2tJI}), this case is usually difficult to tackle analytically. So we prefer not to discuss about this case in the present work, but postpone it into further study instead.
\subsection{Reconstruction from matter-contraction}
\subsubsection{A minimal coupling matter-contraction model}
Now we turn to another possibility that scale-invariance could be obtained by a minimal coupling single scalar field, namely the matter-contraction scenario. In this scenario, the Universe is contracting, and the equation of state of the universe is set to be approximately zero ($w_E\simeq0$), so the scale factor and the Hubble parameter can be parameterized as \be \label{aandHM} a_E=a_\ast\Big(\frac{-t_E}{-t^\ast_E}\Big)^{2/3}~,~H_E=\frac{2}{3t_E}~,\ee and here $t_E$ goes from $-\infty$ to $0_-$. Similar to the inflationary case, we can obtain the solutions of $\varphi_E$ and $V(\varphi_E)$ by making use of Eqs. (\ref{rhoandpE}) and (\ref{friedmannE}), but here we should require that the system has (nearly) no pressure, namely, $(d\varphi_E/dt_E)^2=2V$. The solution that have the scaling behavior is: \bea \label{varphiEM} \varphi_E&=&\frac{2M_{Pl}}{\sqrt{3}}\ln(-M^{-1}t_E)-\frac{2M_{Pl}}{\sqrt{3}}\ln(-M^{-1}t^\ast_E)+\varphi^\ast_E~,\\ V(\varphi_E)&\sim&\frac{2M_{Pl}^2}{3(-t_E)^2}\sim\frac{2}{3}M^2M_{Pl}^2e^{-\sqrt{3}\frac{\varphi_E}{M_{Pl}}}~,\eea where $\varphi^\ast_E=2M_{Pl}\ln(-M^{-1}t^\ast_E)/\sqrt{3}$ can be set to eliminate the integral constants $t^\ast_E$ and $\varphi^\ast_E$. The potential could also be written as \be\label{potentialEM} V(\varphi_E)=V_0e^{-\sqrt{3}\frac{\varphi_E}{M_{Pl}}}~,\ee where $V_0$ is a 4-dimension constant. Note that this solution is consistent with that in the original paper on matter-contraction, by the authors of F. Finelli and R. Brandenberger \cite{Finelli:2001sr}.

\subsubsection{Its nonminimal coupling correspondence}
Now we turn on to reconstructing the form of the action (\ref{actionJ}) which, when transformed to its Einstein frame, can act like the above matter-contracting model and thus lead to scale-invariant power spectrum as well. The approach is exactly the same as was done in Sec. \ref{recinf} for the case of inflation. We parameterize the function $F(\phi)$ as \be\label{parametrizeFM} F(\phi(t_J))=F_\ast\Big(\frac{|t_J|}{|t^\ast_J|}\Big)^{2f_M}~,\ee where $f_M$ is some parameter, and we also have the relation \be \label{relationtM} \int^{t^\ast_E}_{t_E}dt_E=\int^{t^\ast_J}_{t_J}\sqrt{2F}dt_J=\frac{\sqrt{2F_\ast}}{|t^\ast_J|^{f_M}}\int^{t^\ast_J}_{t_J}|t_J|^{f_M}dt_J~\ee from Eq. (\ref{relation}). One can also assume the monotonic increase of $t_E$ and $t_J$, and have two cases of $t_J>0$ and $t_J<0$ respectively, only noticing that $t_E$ is negative here. Since the whole calculation of integrating Eq. (\ref{relationtM}) is a straightforward following of that in Sec. \ref{recinf}, we only give the final result as:  \bea \label{tE2tJM}  -t_E=\left\{ \begin{array}{l} \frac{\sqrt{2F_\ast}(-t^\ast_J)}{f_M+1}\Big(\frac{-t_J}{-t^\ast_J}\Big)^{f_M+1}~~~~{\rm for}~~f_M>-1~,\\\\ \sqrt{2F_\ast}(-t^\ast_J)\ln(-\bar{t}_J)~~~~{\rm for}~~f_M=-1~,\\\\ -\frac{\sqrt{2F_\ast}t^\ast_J}{f_M+1}\Big(\frac{t_J}{t^\ast_J}\Big)^{f_M+1}~~~~{\rm for}~~f_M<-1~,\\ \end{array}\right. \eea where $\bar{t}_J=t_J/t_{Pl}$.

The dependence of $a_J$, $H_J$ and $\phi$ on $t_J$ can also be easily obtained, making use of Eqs. (\ref{aandHM}), (\ref{varphiEM}) as well as (\ref{relation}). First of all, the scale factor in the Jordan frame $a_J$ is: \be a_J(t_J)=\frac{a_E}{\sqrt{2F}}=\frac{a^\ast_E}{(-t^\ast_E)^{2/3}}\frac{(-t_E)^{2/3}}{\sqrt{2F}}~,\ee and making use of the result (\ref{tE2tJM}), we have: \bea \label{a2tJM} a_J(t_J)=\left\{ \begin{array}{l} \frac{a^\ast_E}{(f_M+1)^{2/3}(2F_\ast)^{1/6}}\Big(\frac{-t^\ast_J}{-t^\ast_E}\Big)^{2/3}\Big(\frac{-t_J}{-t^\ast_J}\Big)^{\frac{2-f_M}{3}}~~~~{\rm for}~~f_M>-1~,\\\\ \frac{a^\ast_E(-t^\ast_J)^{-1/3}}{(-t^\ast_E)^{2/3}(2F_\ast)^{1/6}}[\ln(-\bar{t}_J)]^\frac{2}{3}(-t_J)~~~~{\rm for}~~f_M=-1~,\\\\ \frac{a^\ast_E}{(f_M+1)^{2/3}(2F_\ast)^{1/6}}\Big(\frac{t^\ast_J}{t^\ast_E}\Big)^{2/3}\Big(\frac{t_J}{t^\ast_J}\Big)^{\frac{2-f_M}{3}}~~~~{\rm for}~~f_M>-1~,\\ \end{array}\right.\eea respectively. Considering Eq. (\ref{H2tJI}), the Hubble parameter $H_J$ can also be obtained in terms of $t_J$ as: \bea \label{H2tJM} H_J=\left\{ \begin{array}{l} \frac{2-f_M}{3t_J}~~~~{\rm for}~~f_M>-1~{\rm and}~f_M<-1~,\\\\ \frac{1}{t_J}[\frac{2}{3\ln(-\bar{t}_J)}+1]~~~~{\rm for}~~f_M=-1~.\\ \end{array}\right.\eea In the first case, $\epsilon_J$ can also be written as $\epsilon_J=3/(2-f_M)$, so as to have $H_J=1/(\epsilon_Jt_J)$.

\begin{figure}[htbp]
\centering
\includegraphics[scale=0.3]{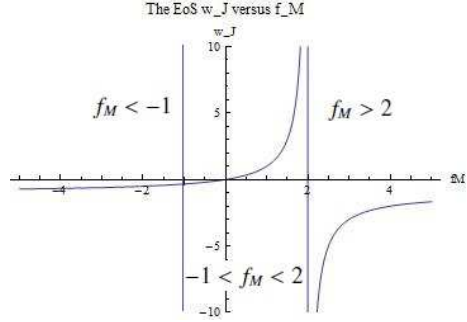}
\caption{The behavior of $w_J$ in $f_M\neq-1$ case w.r.t. $t_J$. The whole space is divided into 3 parts, namely where $f_M<-1$ or $f_M>2$ the universe expands while where $-1<f_M<2$ it contracts. The points of $f_M=2$ corresponds to $w_J\rightarrow\pm\infty$ while $f_M=-1$ corresponds to $w_J=-1/3$. When $f_M=0$, $w_J$ is equal to 0 in the contracting region, which is trivial. The approach of $f_M$ to 2 from left and right hand side leads $w_J$ to positive and negative infinity, which corresponds to slow contraction or expansion, while $f_M\rightarrow\infty$ leads $w_J$ to $-1$ in expanding region, which is inflation.}\label{wJvsfM}
\end{figure}

From Eq. (\ref{H2tJM}), one can find that the two points that divide different behaviors of the universe are $f_M=-1$ and $f_M=2$. When $f_M<-1$ or $f_M>2$, $H_J>0$ which corresponds to an expanding universe in the Jordan frame, while the region where $-1<f_M<2$ leads to a contracting phase. At $f_M=-1$, $H_J$ is also negative and the universe is also contracting. This can also be verified from the viewpoint of $a_J$, that is, $a_J(t_J)$ is increasing for $f_M<-1$ and $f_M>2$, while decreasing for $-1\leq f_M<2$. The equation of state in each case turns out to be: \bea \label{w2tJM} w_J=\left\{ \begin{array}{l} \frac{f_M}{2-f_M}~~~~{\rm for}~~f_M>-1~{\rm and}~f_M<-1~,\\\\ -\frac{\ln(-\bar{t}_J)[8+3\ln(-\bar{t}_J)]}{[2+3\ln(-\bar{t}_J)]^2}~~~~{\rm for}~~f_M=-1~,\\ \end{array}\right.\eea and the relation between $w_J$ and $f_M$ and the region where the universe contracts or expands has been shown in Fig. \ref{wJvsfM}. 

\begin{figure}[htbp]
\centering
\includegraphics[scale=0.5]{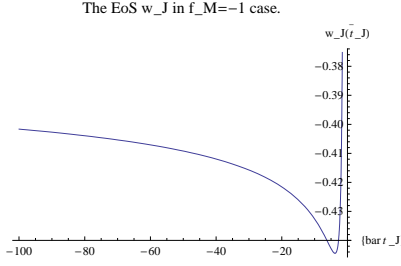}
\caption{The behavior of $w_J$ in $f_M=-1$ case w.r.t. $\bar{t}_J$. In this case, $w_J$ is no longer a constant, especially near $\bar{t}_J=0$, but in the far past when $\bar{t}_J\ll-1$, it will approach to a constant value.}\label{wJMfeq-1}
\end{figure}

As is similar in the inflation case, for the case where $f_M\neq-1$, the condition {\it i)} of being $a_J\sqrt{2F}\sim |\eta_\ast-\eta|^2$ can be satisfied by the matter-contraction background in the Einstein frame with the conformal relation (\ref{relation}), while the condition {\it ii)} can be satisfied by the power-law scalings of $a_J$ and $F$ in terms of $t_J$, as well as some proper $Z(\phi)$. For the $f_M=-1$ case, $w_J$ is no longer a constant, but in the far past when $\bar{t}_J\ll-1$, it will approach to a constant value. As can be seen from the figure, in this case $w_J$ will never go below the bound of $w_J=-1/3$ in the case of contracting universe, while never exceed that bound in the case of expanding phase. The behavior of $w_J$ in $f_M=-1$ case w.r.t. $\bar{t}_J$ is plotted in Fig. \ref{wJMfeq-1}.

Finally let's discuss the dependence of $\phi$ on $t_J$ in order to determine $f_M$ and the equation of state $w_J$, as well as $F(\phi)$ and $U(\phi)$ in terms of $\phi$. The relation between $d\varphi_E/dt_E$ and $\dot\phi$ has already been given in Eq. (\ref{dotphiE2J}) only with $f_I\rightarrow f_M$, but the time-dependence of $d\varphi_E/dt_E$ for matter-contraction follows Eq. (\ref{varphiEM}), which is: \be (\frac{d\varphi_E}{dt_E})^2=\frac{4M_{Pl}^2}{3(-t_E)^2}~,\ee so we have \bea \label{zdotphi2M} Z\dot\phi^2=\left\{ \begin{array}{l} \frac{4M_{Pl}^2F}{3(-t_J)^2}\big[2(f_M+1)^2-9f_M^2\big]~~~~{\rm for}~~f_M>-1~,\\\\ \frac{4M_{Pl}^2F}{3(-t_J)^2}\big[\frac{2}{[\ln(-\bar{t}_J)]^2}-9\big]~~~~{\rm for}~~f_M=-1~,\\\\ \frac{4M_{Pl}^2F}{3t_J^2}\big[2(f_M+1)^2-9f_M^2\big]~~~~{\rm for}~~f_M<-1~\\ \end{array}\right. \eea in matter-contraction case. We also assume that $Z(\phi)=Z_0^M\phi^{2z_M}$, then for $f_M\neq-1$ we have \bea \label{phiJM} \phi^{z_M+1}&=&\frac{(z_M+1)}{f_M}\sqrt{\frac{4M_{Pl}^2F_\ast}{3Z_0^M}\Big[2(f_M+1)^2-9f_M^2\Big]}\big(\frac{\pm t_J}{\pm t_J^\ast}\big)^{f_M}~\nonumber\\ &=&\frac{(z_M+1)}{f_M}\sqrt{\frac{4M_{Pl}^2}{3Z_0^M}\Big[2(f_M+1)^2-9f_M^2\Big]}\sqrt{F}~,\eea and the function $F(\phi)$ can also be presented in terms of $\phi$ as \bea \label{parametrizeF2M}
F&=&F_0^M\phi^{2z_M+2}~,\\ F_0^M&=&\frac{3Z_0^If_M^2}{4M_{Pl}^2(z_M+1)^2}\Big[2(f_M+1)^2-9f_M^2\Big]^{-1}~,\eea which also leads to the fact $c_J$ is constant and $Q_{\cal R}\sim F$. Moreover, from Eq. (\ref{phiJM}), we can express $\phi$ as: \bea \label{parametrizephiJM} \phi&=&\phi_0^M(\pm t_J)^{\frac{f_M}{z_M+1}}~,\\ \phi_0^M&=&\left\{\frac{(z_M+1)}{f_M(\pm t_J^\ast)^{f_M}}\sqrt{\frac{4M_{Pl}^2F_\ast}{3Z_0^M}[2(f_M+1)^2-9f_M^2]}\right\}^\frac{1}{z_M+1}~\eea so that $(\pm t_J)=(\phi/\phi_0^M)^{(z_M+1)/f}$, and the first and second time derivatives of $\phi$ are: \bea \label{parametrizedotphiJM} \dot\phi&=&\pm\frac{f_M}{z_M+1}\phi_0^M(\pm t_J)^\frac{f_M-z_M-1}{z_M+1}~,\\ \ddot\phi&=&\frac{f_M(f_M-z_M-1)}{(z_M+1)^2}\phi_0^M(\pm t_J)^\frac{f_M-2z_M-2}{z_M+1}~,\eea respectively. Finally, from either Eq. (\ref{eomJ}) or Eq. (\ref{friedmannJ}) we have: \bea\label{potentialJM} U(\phi)&=&\frac{4}{3}(1+f_M)^2\frac{F}{(\pm t_J)^2}~\nonumber\\ &=&U_0^M\phi^{2(z_M+1)(1-\frac{1}{f_M})}~,\\ U_0^M&=&\frac{4}{3}(1+f_M)^2 F_0^M(\phi_0^M)^\frac{2(z_M+1)}{f_M}~,\eea which is consistent with the potential in Einstein Frame (\ref{potentialEM}) and the relation (\ref{relation}).

From Eq. (\ref{potentialJM}), we see that for a given $z_M$ and $f_M$, the power-law index of $U(\phi)$ with respect of $\phi$ can be determined, and the evolution behavior of the universe driven by action (\ref{actionJ}) in the Jordan frame can also be determined inversely by the given power-law forms of $U(\phi)$ and $Z(\phi)$. We are more interested in the latter case, so we start with the action (\ref{actionJ}) with \be\label{FandZandUM} F(\phi)=F_0^M\phi^{2z_M+2}~,~~~Z(\phi)=Z_0^M\phi^{2z_M}~,~~~U(\phi)=U_0^M\phi^{q_M}~,\ee which lead to \be\label{qMandfM} q_M=2(z_M+1)(1-\frac{1}{f_M})~,~~~f_M=\frac{2(z_M+1)}{2(z_M+1)-q_M}~.\ee Therefore, with Eq. (\ref{w2tJM}), we have: \be\label{wJM} w_J=\frac{z_M+1}{z_M+1-q_M}~.\ee

Eqs. (\ref{FandZandUM}-\ref{wJM}) are our main result for this subsection, linking the parameters $z_M$ and $q_M$ from the action (\ref{actionJ}) in the Jordan frame, to the behavior of the universe that it may drive, which corresponds to a matter-contraction in the Einstein frame and thus give rise to scale-invariant power spectrum. As is the same for the above subsection, to get scale-invariant power spectrum, the evolution in the Jordan frame also has more degrees of freedom and more action-dependence than that in the Einstein frame. Two examples are also shown in order: one is $z_M=-1/2$ and $q_M=1$, which is the Brans-Dicke-like action where $F(\phi)\sim\phi$, $Z(\phi)\sim\phi^{-1}$ and $U(\phi)\sim\phi$, and in this case we get $w_J\sim-1$, which behaves like inflation. The other is $z_M=0$, $q_M=1+\varepsilon$ where $|\varepsilon|\ll1$. This gives $F(\phi)\sim\phi^2$, $U(\phi)\sim\phi^{1+\varepsilon}$ and $Z(\phi)$ is a constant. This gives $w_J=\pm\infty$, which indicates that the universe is evolving slowly, either expanding or contracting according to the sign of $\varepsilon$.

Due to the difficulty in tackling the $f_M=-1$ case analytically, we will not discuss about it in the present work either, and leave it to further study.

\section{Discussions on Some Other Issues}

In this section, we will extend our discussions on several other Big-Bang puzzles, namely the {\it horizon} and {\it flatness} problems that can be solved both in inflation and matter-contraction scenarios, the {\it singularity} problem that plagues inflation scenario, as well as the {\it anisotropy} problem that plagues matter-contraction scenario. We will see whether/how these problems can exist if we move to another frame, i.e., the Jordan frame, or in other words, whether the conformal transformation could help alleviate these problems, or on the opposite, make them aggravated. We will briefly comment on these problems.
\\
{\it Horizon and flatness problems.} These two problems exist in the Big-Bang scenario but can be avoided both in inflationary \cite{Kolb:1990vq} or matter-contracting \cite{Brandenberger:2009jq} scenarios. We have already shown in the previous section that the counterparts of both the two cases will not suffer from the horizon problem. As can be seen on Figs. \ref{wJvsfI} and \ref{wJvsfM}, both inflationary and matter-contracting scenarios in Einstein frame only correspond to expanding universe with $w_J<-1/3$ or contracting universe with $w_J>-1/3$ in Jordan frame, which would not lead to the problem of inhomogeneity, which is an important observation in this paper. This issue can also be addressed from another point of view, say, the evolution of perturbations. If we go backwards, we need to have the horizon grow faster than the wavelengths of fluctuation modes, in order to set all the fluctuation modes inside the horizon where it is causal connected. In the Einstein frame, the horizon has the same scaling of $(a_E\sqrt{2\epsilon_E})^{\prime\prime}/(a_E\sqrt{2\epsilon_E})$ for curvature perturbations and $a_E^{\prime\prime}/a_E$ for tensor perturbations, while when transformed into the Jordan frame, from the relations (\ref{relation}) and (\ref{relation2}) we know that it will remain invariant for both curvature and tensor perturbations, so we will not have horizon problem in our nonminimal coupling theory. The flatness problem will also do no harm to us since the scaling of the deviation of the spatial curvature $\Omega_K$ from 0 is as $(aH)^{-2}$, which is also invariant under conformal transformation. These are consistent with the equivalence of the two frames.
\\
{\it Singularity problem.} The singularity problem has been stressed in the Singularity Theorems \cite{Hawking:1973uf,Borde:1993xh,Novello:2008ra} that in the inflation scenario when we go {\it backwards}, the singularity must be reached where the scale factor $a(t)$ shrinks to zero at some finite time, as long as the Hubble parameter $H$ keeps on larger than 0. In order to prevent the singularity, we should let $H$ reach or go below 0 to stop $a(t)$ from shrinking, either becoming static or expanding (which is the picture of bouncing cosmologies \cite{Cai:2007qw,Novello:2008ra}). That requires a period when $\dot H>0$, or super-inflation \cite{Gunzig:2000kk}, which couldn't be realized by a single canonical scalar field minimal coupled to gravity. \footnote{For minimal coupling theory, $\dot H>0$ may be realized by the phantom inflation model \cite{Piao:2003ty,Piao:2004tq}, however, how to remove the ghost or quantum instability still remains a challenging issue \cite{Carroll:2003st}. See comments in e.g. \cite{Liu:2012ib}. } However, as for the nonminimal coupling theories, we could obtain the $\dot H>0$ region in its Jordan frame due to the coupling to gravity. For instance, in the cases that we are considering, it could be the region where $H_J$ is given by Eq. (\ref{H2tJI}) with $f_I<1/(\epsilon_E-1)$. In that region, $\dot H_J=-[f_I(1-\epsilon_E)+1]/(\epsilon_E t_J^2)>0$, and one can also see from Fig. \ref{wJvsfI} that $w_J$ is less than $-1$. It indicates that one may have a non-singular scenario with a nonminimal coupling theory that would cause super-inflation \cite{Gunzig:2000kk} or even bounce \cite{Cai:2007qw,Setare:2008qr,Qiu:2010vk}. Furthermore, since it is conformally equivalent to the normal inflationary scenario, the perturbations will be ghost-free and the scale-invariant power-spectrum is available.
\\
{\it Anisotropy problem.} This is a notorious problem that generally exists in matter-contracting as well as bouncing models \cite{Kunze:1999xp}. In GR, if we allow the existence of even a tiny amount of anisotropy at the very beginning of the universe, the anisotropy will evolve with scaling of $a^{-6}(t)$ where $a(t)$ is the scale factor of the universe. In inflationary phase things will be fine, since it will decay fast when $a(t)$ becomes larger and larger. However in the contracting phase where $a(t)$ is getting smaller, that will cause a problem. The anisotropy will grow fast and dominate the universe, which will make it collapse to a totally anisotropic one. In order to avoid the domination of the anisotropy, one has to expect a contracting background that evolves even faster, which requires an equation of state larger than unity such as the Ekpyrotic scenario \cite{Khoury:2001wf}. Since by conformal transformation we can get a different equation of state in a different frame, it ignites our curiosity about whether this problem can also be get rid of naively in this way. However, careful check has to be done. To be precise, let's consider the Bianchi-IX metric \cite{Misner:1974qy} as follows: \be ds^2=-dt^2+a^2(t)\sum_{i=1}^3e^{2\beta_i(t)}d{x^i}^2~,~~~~{\rm with}~~\beta_1(t)+\beta_2(t)+\beta_3(t)=0~.\ee Then the Friedmann Equation in the Jordan frame (\ref{friedmannJ}) will be modified as: \be\label{friedmannani} 3M_{Pl}^2H_J^2-\frac{1}{2}M_{Pl}^2(\sum_{i=1}^3\dot\beta_i^2)=\rho_J~,\ee where $\rho_J$ is defined in Eq. (\ref{rhoJ}), the second term in the left hand side can be defined as the anisotropy term $\sigma^2$, and the $\beta_i$'s satisfy the equation \be \ddot\beta_i+(3H_J+\frac{\dot F}{F})\dot\beta_i=0~.\ee

Since the anisotropy problem only exists in contraction scenarios, we only focus on the contraction scenarios driven by the nonminimal coupling theory in the Jordan frame. It splits into two cases, one is the case which is reconstructed from inflation in the Einstein frame, with $1/(\epsilon_E-1)<f_I<-1$, and the other is the case which is reconstructed from matter-contraction in the Einstein frame, with $-1\leq f_M<2$. In the first case, from the expressions of $H_J$ and $F$ in terms of $t_J$, namely Eqs. (\ref{H2tJI}) and (\ref{parametrizeFI}), we have: \be \ddot\beta_i+\frac{3(f_I+1)-f_I\epsilon_J}{\epsilon_J t_J}\dot\beta_i=0~.\ee This gives: \be \dot\beta\sim(-t_J)^{\frac{f_I\epsilon_J-3(f_I+1)}{\epsilon_J}}~,\ee and therefore the anisotropy term defined in Eq. (\ref{friedmannani}) (in the Jordan frame) evolves as: \be \sigma^2=\frac{1}{2}\sum_i\dot\beta_i^2\sim(-t_J)^{\frac{2f_I\epsilon_J-6(f_I+1)}{\epsilon_J}}~.\ee

On the other hand, from Eq. (\ref{rhoJ}) as well as Eqs. (\ref{H2tJI}), (\ref{parametrizephiJI}), (\ref{parametrizedotphiJI}), (\ref{FandZandUI}) and (\ref{qIandfI}) we know that the effective energy density for the Jordan frame background evolves as $\rho_J\sim(-t_J)^{-2}$, therefore in order not to have anisotropy term exceed the background evolution, we need: \be \frac{2f_I\epsilon_J-6(f_I+1)}{\epsilon_J}>-2\Rightarrow \epsilon_E<3~,~\frac{1}{\epsilon_E-1}<f_I<-1~.\ee This gives the condition for our theory to be free of anisotropy problem. This totally coincides with the condition in the Einstein frame, that is $w_E<1$, and in our case where $\epsilon_E\ll 1$, it satisfy the condition very well. This is not surprising, since the two frames are equivalent, and the problem should not appear in one frame while disappear in the other. However what is interesting is, it provides an example that one might build a contracting scenario free of anisotropy problem with scale-invariant power spectrum. This case may deserve further investigations in the future.

However, in the second case one gets: \bea \ddot\beta_i+\frac{2+f_M}{t_J}\dot\beta_i&=&0~~~~{\rm for}~~-1<f_M<2~,\\ \ddot\beta_i+\frac{1}{t_J}[\frac{2}{\ln(-\bar{t}_J)}+1]\dot\beta_i&=&0~~~~{\rm for}~~f_M=-1~,\eea where the expressions of $H_J$ and $F$ in terms of $t_J$ are given by Eqs. (\ref{H2tJM}) and (\ref{parametrizeFM}). The equation for $-1<f_M<2$ case gives: \be \dot\beta\sim(-t_J)^{-(2+f_M)}~,\ee and therefore the anisotropy term evolves as $\sigma^2\sim(-t_J)^{-2(2+f_M)}$. On the other hand, the effective energy density defined in Eq. (\ref{rhoJ}) evolves as $\rho_J\sim(-t_J)^{-2}$, where we have made use of Eqs. (\ref{H2tJM}), (\ref{parametrizephiJM}), (\ref{parametrizedotphiJM}), (\ref{FandZandUM}) as well as (\ref{qMandfM}). Therefore the requirement for the anisotropy term not to exceed the background evolution is: \be -2(2+f_M)>-2\Rightarrow f_M<-1~,\ee which contradicts with our condition of $-1<f_M<2$.

Similarly, the equation for $f_M=-1$ indicates: \be \dot\beta\sim\frac{1}{(-t_J)[\ln(-\bar{t}_J)]^{2}}~,\ee which denotes the scaling of anisotropy term as $\sigma^2\sim(-t_J)^{-2}[\ln(-\bar{t}_J)]^{-4}$. Since the factor $[\ln(-\bar{t}_J)]^{-4}$ is monotonically growing, the anisotropy definitely grows faster than the background, therefore it is impossible to get an anisotropy-free contracting phase for the nonminimal coupling theory constructed from a matter-contracting phase. As mentioned above, this is not surprising due to the equivalence between the two frames. However, there also has been some other attempts to get rid of this problem in matter-contracting scenario, including having matter with non-linear equation of state, or ghost condensate, see \cite{Bozza:2009jx}.
\section{Conclusion}
Nonminimal coupling theories with a single scalar field such as $F(\phi)R$ can be transformed to its Einstein frame via an appropriate conformal transformation (\ref{conformal}), where it behaves as a minimal coupling theory in normal Einstein's gravity. The two theories before and after transformation are equivalent, rendering the same observational results. Making use of this property, one may either start with a nonminimal coupling theory in the Jordan frame and move to the Einstein frame to simplify the calculations, or start from a minimal coupling theory, which can be viewed as a Einstein frame presentation, to find what the theory is like in the Jordan frame. However, it is non-trivial to perform the latter, since we do not know {\it a priori} the form of $\Omega^2$ in Eq. (\ref{conformal}), so one may therefore need to use the reconstruction method.

In this paper, we focus on reconstructing nonminimal coupling theories with two minimal coupling cases which can give rise to scale-invariant power spectrum, namely inflation and matter-contraction, and obtained the relation between the equation of state of the universe in the Jordan frame and the parameters in the nonminimal coupling action. Therefore, as long as we set appropriate parameters to the action and let it evolve as expected, we can get a scale-invariant power spectrum which fits the observational data. We found that both scenarios can correspond to expanding and contracting scenarios in the Jordan frame, but due to the small slow-roll parameter in the inflation scenario, there is only a narrow parameter space for the Jordan frame theory to be in a contracting phase. Moreover, we checked the solutions of the various Big-Bang puzzles in both frames, such as the horizon problem, the flatness problem, the singularity problem, as well as the anisotropy problem. Due to the equivalence of the two frames, one may be able to find models that can be free of these problems.

Before ending this paper, let's remark that the same approach of reconstruction can also be used to reconstruct $f(R)$ theories, which is also equivalent to a normal scalar field theory via conformal transformation. In Ref. \cite{Nojiri:2006be}, the authors have considered the reconstruction of $f(R)$ theories, but only in expanding backgrounds. Nonetheless, if we start with a scalar field that can drive inflation or matter-contraction, then by such reconstruction, we can also obtain theories with scale-invariant power spectrum, which can describe our universe within pure (but extended) gravity theory, instead of invoking extra scalar fields. This work will be done in a coming paper \cite{preparation}.

\acknowledgments
The author thanks Pisin Chen, Yun-Song Piao, Robert Brandenberger, Antonio De Felice and Je-An Gu for their useful suggestions during discussions, and also the anonymous referee for his nice suggestions, according to which the paper has been improved. He also enjoys the ``2012 Asia Pacific School/Workshop on Cosmology and Gravitation" held in Yukawa Institute for Theoretical Physics in Kyoto University, and the visit in Research Center for the Early Universe and the Kavli Institute for the Physics and Mathematics of the Universe in Tokyo University, Japan, when this work is finalized. This work is funded in part by the National Science Council of R.O.C. under Grant No. NSC99-2112-M-033-005-MY3 and No. NSC99-2811-M-033-008 and by the National Center for Theoretical Sciences.

\end{document}